\documentclass{PoS}

\usepackage{amsmath}

\newcommand{\mcl}[1]{\multicolumn{5}{c|}{ {#1} }}
\newcommand{\mc}[1]{\multicolumn{5}{c}{ {#1} }}
\newcommand{\crl}[1]{\multicolumn{1}{r|}{ {#1} }}
\newcommand{\crr}[1]{\multicolumn{1}{r}{ {#1} }}

\title{2+1 flavour thermal studies on an anisotropic lattice}

\ShortTitle{2+1 flavour thermal studies on an anisotropic lattice}

\author{\speaker{Chris Allton}$^{,a,\dagger}$,
Gert Aarts$^a$,
Alessandro Amato$^{a,b}$,
Wynne Evans$^a$,
Pietro Giudice$^c$,
Timothy Harris$^d$,
Simon Hands$^a$,
Aoife Kelly$^e$,
Sin\'ead M. Ryan$^d$
and
Jon-Ivar Skullerud$^e$
\\

$^a$  Department of Physics, College of Science, 
  Swansea University, Swansea, United Kingdom\\

$^b$  Institut f\"ur Theoretische Physik, Universit\"at Regensburg, 
  Regensburg, Germany\\

$^c$  Universit\"at M\"unster, Institut f\"ur Theoretische Physik,
  M\"unster, Germany\\

$^d$  School of Mathematics, Trinity College, Dublin 2, Ireland\\

$^e$  Department of Mathematical Physics, National University of Ireland Maynooth, 
  Maynooth, County Kildare, Ireland\\

$^\dagger$ E-mail: c.allton@swan.ac.uk

}

\abstract{The {\sc fastsum} collaboration has initiated a detailed
  study of thermal QCD using 2+1 flavours of improved Wilson quarks on
  anisotropic lattices. Spatial volumes of $(3\mbox{fm})^3$ and
  $(4\mbox{fm})^3$ are used at fixed cut-off with temperatures ranging
  from 40 to 350 MeV (corresponding to temporal lattice extents of 128
  to 16 lattice units). Results presented here include the
  deconfinement temperature and a study of the restoration of chiral
  symmetry, together with a brief summary of our collaboration's other
  results from these ensembles.}

\FullConference{31st International Symposium on Lattice Field Theory
  LATTICE 2013\\
   July 29  August 3, 2013\\
    Mainz, Germany}

\begin{document}

%}}}

%{{{ Introduction

\section{Introduction}
Particle physics data are famously collated and summarised in the
Particle Data Book \cite{Beringer:1900zz}. However, it is interesting
to note that there are no entries on the deconfined phase of QCD -- a
symptom of the difficulty of studying (experimentally and
theoretically) this new phase. The {\sc fastsum} Collaboration has
studied QCD at non-zero temperature for a number of years using
dynamical quarks on {\em anisotropic} lattices where the temporal
lattice spacing, $a_\tau$, is less than the spatial one, $a_s$. Since
the temperature $T = 1/(a_\tau N_\tau)$, where $N_\tau$ is the number
of lattice points in the temporal direction, this gives the distinct
advantage that more points are sampled in a euclidean correlator for a
given temperature compared to the isotropic case.

Our research programme began with two-flavour dynamical ``1st
generation'' ensembles from which we studied a number of
phenomenological quantities, such as spectral features in charmonium
and bottomonium at zero and non-zero momenta, and the inter-quark
potential in charmonium \cite{fastsum1}. We have now improved the
accuracy of our results by producing our ``2nd generation'' lattices
which have 2+1 dynamical flavours, a larger volume, improved discretisation
and more realistic dynamical quark masses, see Table
\ref{tab:params}.

In this talk, I give an overview of our 2nd generation ensembles
including our estimate of the deconfinement temperature, $T_c$, obtained
from the Polyakov loop. I discuss the partial restoration of chiral
symmetry in the light meson spectrum and briefly review four results
obtained from these lattices which are covered fully in other talks
\cite{Giudice:2013fza,Amato:2013oja,Evans:2013zca,Harris:2013vfa}.

%}}}

%{{{ Lattice details

\section{Lattice details}

Our 2nd generation ensembles use the Hadron Spectrum Collaboration's
(HSC) Symanzik-improved gauge action \cite{Lin:2008pr}, with
\begin{equation}
S_G = \frac{\beta}{\gamma_g} \left\{
\sum_{x,s\neq s^\prime} \left[
    \frac{5}{6 u_s^4     }{\cal P}_{ss^\prime}(x)
  - \frac{1}{12 u_s^6    }{\cal R}_{ss^\prime}(x)\right]
+ \sum_{x,s}\gamma_g^2 \left[
    \frac{4}{ 3 u_s^2 u_\tau^2}{\cal P}_{s\tau}(x)
  - \frac{1}{12 u_s^4 u_\tau^2}{\cal R}_{s\tau}(x)\right]
\right\},
\label{eq:sg}
\end{equation}
where ${\cal P}$ and ${\cal R}$ are the usual $1\times1$ plaquette and
$2\times1$ rectangular Wilson loops, $u_{s(\tau)}$ are the spatial
(temporal) tadpole factors of the bare links, $\gamma_{g(f)}$ are the
bare gauge (fermion) anisotropies and, as usual, $\beta=2 N_c/g^2$ and
$N_c=3$ is the number of colours.  The means of the
stout-smeared links are $\tilde{u}_\mu$ (with $\tilde{u}_\tau=1$).

We use a tadpole-improved clover fermion action and stout-smeared
links \cite{Morningstar:2003gk} using the same parameters as the
Hadron Spectrum Collaboration \cite{Lin:2008pr},
\begin{eqnarray}
S_F
&=& \sum_{x} \overline{\psi }(x) \frac{1}{ \tilde{u}_\tau} \bigg\{ %\left\{
 \tilde{u}_\tau m_0
% wilson terms
                          +  \gamma_\tau \nabla_\tau + \nabla_\tau^2
 +\frac{1}{\gamma_f} \sum_s[ \gamma_s    \nabla_s    + \nabla_s^2] \nonumber \\
% clover terms
 &-&\frac{1}{2} c_\tau \sum_{s}          \sigma_{\tau s}    F_{\tau s}
  - \frac{1}{2} c_s    \sum_{s<s^\prime} \sigma_{ss^\prime} F_{ss^\prime}
%      \right\}
\bigg\} \psi (x),
\label{eq:sf}
\end{eqnarray}
where
\begin{equation}
c_\tau = \left(\frac{\gamma_g}{\gamma_f}+\frac{1}{\xi}\right)\frac{1}{2\tilde{u}_s^2}\;,
\;\;\;\;\;\;\;\;\;\;
c_s    = \frac{1}{\gamma_f \tilde{u}_s^3}.
\end{equation}
The first line is the usual Wilson action and the second line is the
clover piece with $\tau$ and $s$ referring to temporal and spatial
directions. The $\nabla_\mu$ are covariant finite differences and $\xi
= a_s/a_\tau$ is the renormalised anisotropy.  $\gamma_{s(\tau)}$ are
the spatial (temporal) Dirac matrices and $\sigma_{\mu\nu} =
\frac{1}{2}[\gamma_\mu, \gamma_\nu]$.

We use the same parameters as the HSC employed in their studies
\cite{Edwards:2008ja} corresponding to an anisotropy, $\xi=3.5$. We
generate ensembles with two volumes, $24^3$ and $32^3$, enabling us to
study finite volume effects. We also make use of the $T=0$
(i.e. $N_\tau=128$) configurations kindly made available to us from
HSC.  Table \ref{tab:params} gives a full list of our parameters.  The
generation of the ensembles were performed using the Chroma software
suite \cite{Edwards:2004sx} with Bagel routines \cite{bagel}.

%{{{ tab:latt

\begin{table}
\begin{center}
\begin{tabular}{l|ccccc|ccccc}
\hline
&&&&&&&&&&\\
                  & \mcl{\bf 1st Generation} & \mc{\bf 2nd Generation} \\
&&&&&&&&&&\\
\hline
&&&&&&&&&&\\
Flavours          & \mcl{2}                  & \mc{2+1} \\
Volume(s)         & \mcl{($\sim$2fm)$^3$}    & \mc{($\sim$3fm)$^3$ \& ($\sim$4fm)$^3$} \\
$a_\tau$ [fm]     & \mcl{0.0268(1)}          & \mc{0.03506(23)} \\
$a_s$    [fm]     & \mcl{0.162(4)}           & \mc{0.1227(8)} \\
$\xi=a_s/a_\tau$  & \mcl{6}                  & \mc{3.5} \\
$M_\pi/M_\rho$    & \mcl{$\sim 0.54$}        & \mc{$\sim 0.45$} \\
$N_\tau^{\text{crit}} = (a_\tau T_c)^{-1}$
                  & \mcl{33.5}               & \mc{30.4(7)} \\
&&&&&&&&&&\\
Gauge Action      & \mcl{Symanzik Improved}  & \mc{Symanzik Improved} \\
Fermion Action    & \mcl{Stout Link, Fine-Wilson,}
                                             & \mc{Tadpole Improved Clover} \\
                  & \mcl{Coarse-Hamber-Wu}   & \mc{\ } \\
&&&&&&&&&&\\
\hline
&&&&&&&&&&\\
&$N_s$ &$N_\tau$ &$T$ &$T/T_c$ &$N_\text{cfg}$ &$N_s$ &$N_\tau$ &$T$ &$T/T_c$ &$N_\text{cfg}$ \\
&&& (MeV) &&&&& (MeV) &&\\
&&&&&&&&&&\\
\hline 
&&&&&&&&&&\\
                  & 12 & 16 & 459 & 2.09 & \crl{1000} & 24 & 16 & 352 & 1.90 & \crr{1000}\\
                  & 12 & 18 & 408 & 1.86 & \crl{1000} & 32 & 16 & 352 & 1.90 & \crr{1000}\\
                  & 12 & 20 & 368 & 1.68 & \crl{1000} & 24 & 20 & 281 & 1.52 & \crr{1000}\\
                  & 12 & 24 & 306 & 1.40 & \crl{ 500} & 24 & 24 & 235 & 1.27 & \crr{1000}\\
                  & 12 & 28 & 263 & 1.20 & \crl{1000} & 32 & 24 & 235 & 1.27 & \crr{ 500}\\
                  & 12 & 32 & 230 & 1.05 & \crl{1000} & 24 & 28 & 201 & 1.09 & \crr{1000}\\
                  & 12 & 80 &  92 & 0.42 & \crl{ 250} & 32 & 28 & 201 & 1.09 & \crr{ 500}\\
                  &    &    &     &      &            & 24 & 32 & 176 & 0.95 & \crr{1000}\\
                  &    &    &     &      &            & 32 & 32 & 176 & 0.95 & \crr{ 500}\\
                  &    &    &     &      &            & 24 & 36 & 156 & 0.84 & \crr{ 500}\\
                  &    &    &     &      &            & 24 & 40 & 141 & 0.76 & \crr{ 500}\\
                  &    &    &     &      &            & 32 & 48 & 117 & 0.63 & \crr{ 250}\\
                  &    &    &     &      &            & 16 &128 &  44 & 0.24 & \crr{ 500}\\
                  &    &    &     &      &            & 24 &128 &  44 & 0.24 & \crr{ 550}\\
&&&&&&&&&&\\
\hline
\end{tabular}
\caption{A list of the lattice parameters used for our
  1st and 2nd generation ensembles.
\label{tab:params}}
\end{center}
\end{table}

%}}}

%}}}

%{{{ Determination of the deconfining temperature

\section{Determination of the deconfining temperature}

The Polyakov loop, $L$, can be used to determine $T_c$ as follows
\cite{Borsanyi:2012xf}. We note that $L$ is related to the free energy,
$F$, of a static quark, via:
\begin{equation}
L(T) = e^{-F(T)/T}.
\end{equation}
However, $F$ is only defined up to an additive renormalisation
constant, $\Delta F=f(\beta,m_0)$.
We can impose a renormalisation condition at a renormalisation
temperature, $T_R$, by requiring
\begin{equation}
L_R(T_R) \equiv c,
\label{eq:lr}
\end{equation}
for some suitable choice of $T_R$ and $c$. This means a
multiplicative renormalisation constant, $Z_L$, can be fixed as
follows
\begin{equation}
L_R(T) = e^{-F_R(T)/T} = e^{-(F_0(T)+\Delta F)/T} =
L_0(T) e^{-\Delta F/T} = L_0(T) Z_L^{N_\tau}.
\end{equation}

\begin{figure}
\begin{center}
\includegraphics[width=0.9\textwidth]{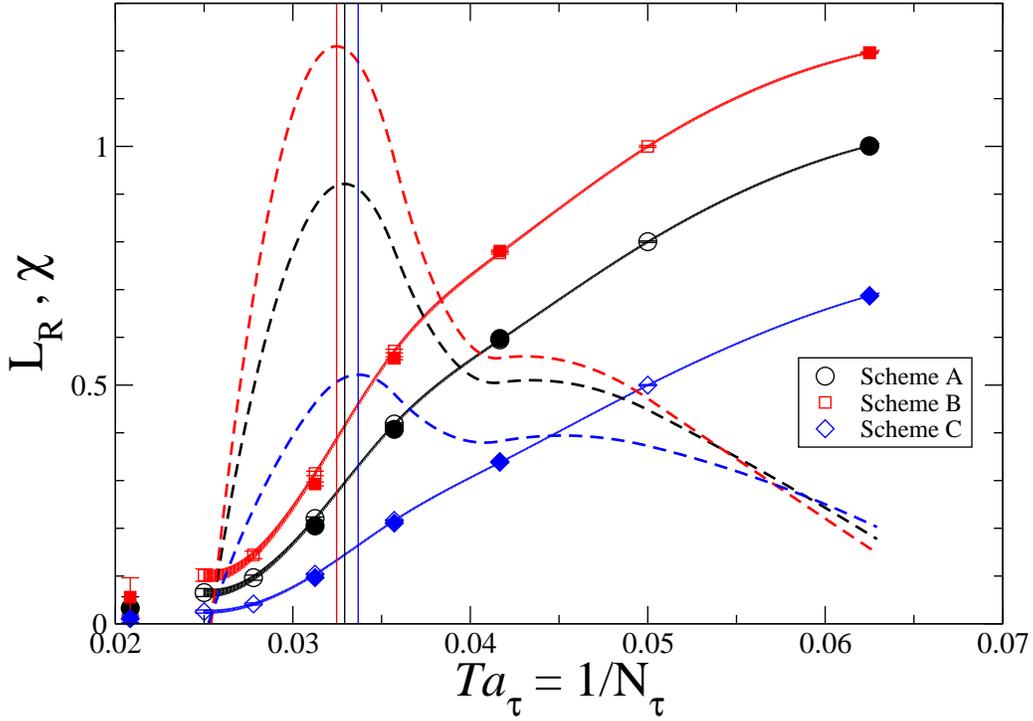}
\caption{The renormalised Polyakov loop, $L_R$, depicted by solid
  ($32^3$) and open ($24^3$) symbols. The solid curves are obtained by
  cubic splines and their temperature derivatives, $\chi$, are
  depicted by dashed curves. Three renormalisation schemes are
  considered, Scheme A: $L_R(N_\tau=16) = 1.0$, Scheme B:
  $L_R(N_\tau=20) = 1.0$, Scheme C: $L_R(N_\tau=20) = 0.5$.}
\label{fig:poly}
\end{center}
\end{figure}

In Fig. \ref{fig:poly}, we plot the Polyakov loop with three different
renormalisation schemes corresponding to different choices of $T_R$
and the constant in Eq.$\;$(\ref{eq:lr}), as listed in the figure
caption. By fitting the data to cubic splines we obtain the point of
inflection $a_\tau T_c = 0.0329(7)$ where the error reflects the
spread from the three renormalisation schemes.  This statistical
uncertainty is given by the thickness of the three interpolating
curves and can be seen to be negligible in this context. The result is
then $N_\tau^\text{crit} = 30.4(7)$ or $T_c = 185(4)$ MeV.

%}}}

%{{{ First results

\section{First results}

\begin{figure}
\begin{center}
\includegraphics[width=0.9\textwidth]{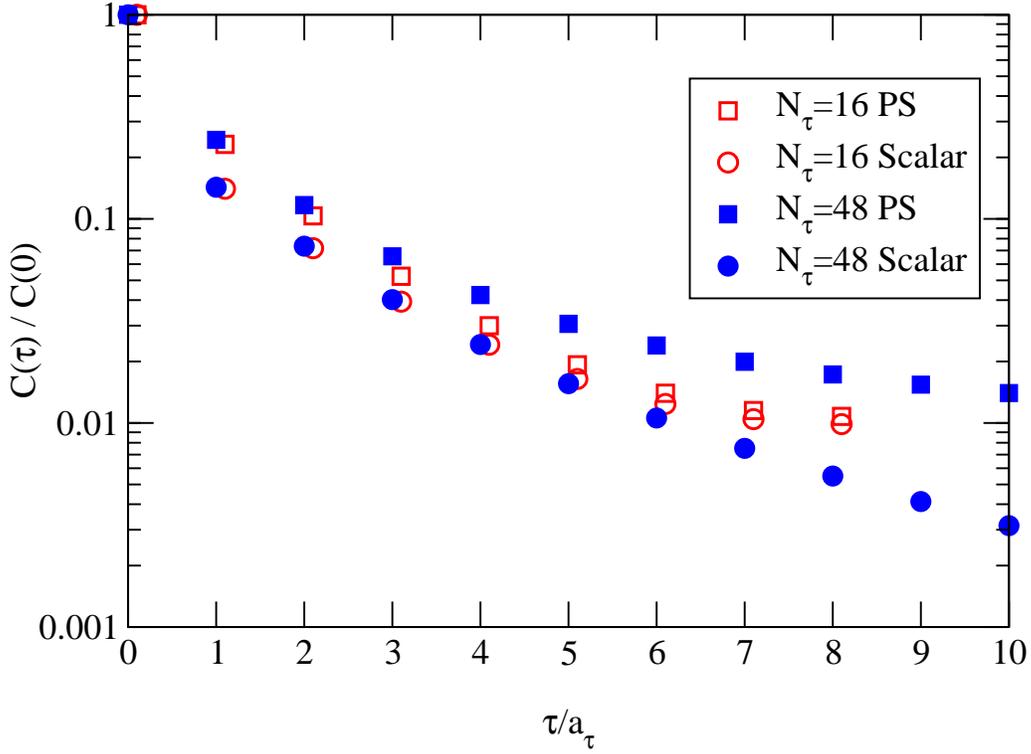}
\caption{Correlation functions (normalised relative to $\tau=0$) for
  the light scalar and pseudoscalar mesons at two different
  temperatures on either side of the deconfinement transition, showing
  partial restoration of chiral symmetry. The $N_\tau=16$ points have
  been shifted horizontally for clarity.}
\label{fig:chiral}
\end{center}
\end{figure}

The deconfinement transition is expected to occur in the same $T$
range as the chiral symmetry restoration. For this reason it is
interesting to study the chiral partners in the light meson sector to
find evidence of this effect. In Fig. \ref{fig:chiral} we show the
pseudoscalar and scalar meson correlator for $T/T_c = 0.63$ and
$1.90$. As can be seen, in the high temperature case, these two
channels are closer together than at low temperature illustrating the
partial restoration of chiral symmetry in these quantities.

We have commenced studying several quantities on our 2nd generation
ensembles. Results on the following quantities have been reported
in this conference and elsewhere.
\begin{itemize}
\item {\bf Susceptibility} \cite{Giudice:2013fza}. We study the
  electric charge susceptibility which is of interest experimentally
  to quantify fluctuations in heavy-ion collision experiments and for
  the determination of the electric charge diffusion coefficient.

\item {\bf Electrical conductivity}
  \cite{Amato:2013oja,Amato:2013naa}. The temperature dependence of
  the electrical conductivity has been calculated on our lattices,
  using the exactly conserved lattice current. We find that the
  conductivity divided by the temperature increases with temperature
  across the deconfinement transition. This is the first time this
  quantity has been computed as a function of temperature.

\item {\bf Inter-quark potential in charmonium} \cite{Evans:2013zca}.
  This is the first time this quantity has been calculated at high
  temperature with relativistic (rather than static) quarks. We find
  that its behaviour at low temperature agrees with the (confining)
  Cornell potential and that it becomes less confining as the
  temperature increases.

\item {\bf Bottomonium spectrum} \cite{Harris:2013vfa}. We have
  used the NRQCD formulation to study spectral functions in
  bottomonium via the Maximum Entropy Method. We confirm our earlier
  result \cite{fastsum1} that the S-wave ($\Upsilon$ and $\eta_b$)
  ground states survive to $T \sim 2T_c$ whereas excited states are
  suppressed, while the P-wave ($h_b, \chi_{b0,b1,b2}$) ground states
  dissociate close to $T_c$.

\item {\bf Charmonium spectrum} \cite{Kelly:sewm2012}. A study of
  charmonium spectral functions across the deconfining transition is
  also in progress.

\end{itemize}

%}}}

%{{{ Conclusions

\newpage\section{Conclusions}

This talk summarises our {\sc fastsum} collaboration's latest
finite-temperature studies using anisotropic lattices. We have
improved upon our 1st generation 2-flavour ensembles by generating
ensembles which have 2+1 flavours, larger volume, improved
discretisation, and smaller dynamical quark masses. In this talk, the
deconfining temperature was presented and the (partial) chiral symmetry
restoration in the light meson sector was studied. Other work presented elsewhere in this
conference was summarised: the susceptibility, electrical
conductivity, interquark potential in charmonium and (NRQCD)
bottomonium spectral functions.

Our future plans are to improve our ensembles further -- we are
currently tuning our ``3rd generation'' ensembles which have a smaller
temporal lattice spacing and have plans for a ``4th generation'' run
with smaller spatial lattice spacing. We will thus be able to move
towards a continuum extrapolation of all our quantities, leading to
truly quantitative finite-temperature results for spectral quantities.

%}}}

%{{{ Acknowledgements

\section*{Acknowledgements}

This work is undertaken as part of the UKQCD collaboration
and the DiRAC Facility jointly funded by STFC, the Large Facilities Capital
Fund of BIS and Swansea University.
We acknowledge the PRACE Grants 2011040469 and Pra05\_1129,
European Union Grant Agreement No. 238353 (ITN STRONGnet),
HPC Wales,
the Irish Centre for High-End Computing,
the Irish Research Council,
the Leverhulme Trust,
the Royal Society,
the Science Foundation Ireland,
STFC,
and the Wolfson Foundation
for support.
The authors would like to thank Seyong Kim, Maria Paola Lombardo, Mike
Peardon and Don Sinclair, for useful comments, discussions and collaboration.

%}}}

%{{{ bibliography

%}}}


\begin{thebibliography}{99}


%\cite{Beringer:1900zz}
\bibitem{Beringer:1900zz}
  J.~Beringer {\it et al.}  [Particle Data Group Collaboration],
  %``Review of Particle Physics (RPP),''
  Phys.\ Rev.\ D {\bf 86} (2012) 010001.
  %%CITATION = PHRVA,D86,010001;%%

\bibitem{fastsum1}
%\cite{Aarts:2007pk}
%\bibitem{Aarts:2007pk}
  G.~Aarts, C.~Allton, M.~B.~Oktay, M.~Peardon and J.~-I.~Skullerud,
  %``Charmonium at high temperature in two-flavor QCD,''
  Phys.\ Rev.\ D {\bf 76} (2007) 094513
  [arXiv:0705.2198 [hep-lat]],
  %%CITATION = ARXIV:0705.2198;%%
  %137 citations counted in INSPIRE as of 14 Nov 2013
%\cite{Oktay:2010tf}
%\bibitem{Oktay:2010tf}
  M.~B.~Oktay and J.~-I.~Skullerud,
  %``Momentum-dependence of charmonium spectral functions from lattice
  %QCD,''
  arXiv:1005.1209 [hep-lat],
  %%CITATION = ARXIV:1005.1209;%%
  %8 citations counted in INSPIRE as of 14 Nov 2013
%\cite{Aarts:2010ek}
%\bibitem{Aarts:2010ek}
  G.~Aarts, S.~Kim, M.~P.~Lombardo, M.~B.~Oktay, S.~M.~Ryan,
  D.~K.~Sinclair and J.~-I.~Skullerud,
  %``Bottomonium above deconfinement in lattice nonrelativistic QCD,''
  Phys.\ Rev.\ Lett.\  {\bf 106} (2011) 061602
  [arXiv:1010.3725 [hep-lat]],
  %%CITATION = ARXIV:1010.3725;%%
  %38 citations counted in INSPIRE as of 14 Nov 2013
%\cite{Aarts:2011sm}
%\bibitem{Aarts:2011sm}
  G.~Aarts, C.~Allton, S.~Kim, M.~P.~Lombardo, M.~B.~Oktay,
  S.~M.~Ryan, D.~K.~Sinclair and J.~I.~Skullerud,
  %``What happens to the Upsilon and eta_b in the quark-gluon plasma?
  %Bottomonium spectral functions from lattice QCD,''
  JHEP {\bf 1111} (2011) 103
  [arXiv:1109.4496 [hep-lat]],
  %%CITATION = ARXIV:1109.4496;%%
  %43 citations counted in INSPIRE as of 14 Nov 2013
%\cite{Aarts:2012ka}
%\bibitem{Aarts:2012ka}
  G.~Aarts, C.~Allton, S.~Kim, M.~P.~Lombardo, M.~B.~Oktay,
  S.~M.~Ryan, D.~K.~Sinclair and J.~-I.~Skullerud,
  %``S wave bottomonium states moving in a quark-gluon plasma from
  %lattice NRQCD,''
  JHEP {\bf 1303} (2013) 084
  [arXiv:1210.2903 [hep-lat]],
  %%CITATION = ARXIV:1210.2903;%%
  %14 citations counted in INSPIRE as of 14 Nov 2013
%\cite{Evans:2013yva}
%\bibitem{Evans:2013yva}
  P.~W.~M.~Evans, C.~R.~Allton and J.~-I.~Skullerud,
  %``Ab Initio Calculation of Finite Temperature Charmonium
  %Potentials,''
  arXiv:1303.5331 [hep-lat],
  %%CITATION = ARXIV:1303.5331;%%
  %3 citations counted in INSPIRE as of 14 Nov 2013
%\cite{Aarts:2013kaa}
%\bibitem{Aarts:2013kaa}
  G.~Aarts, C.~Allton, S.~Kim, M.~P.~Lombardo, S.~M.~Ryan and
  J.~-I.~Skullerud,
  %``Melting of P wave bottomonium states in the quark-gluon plasma
  %from lattice NRQCD,''
  arXiv:1310.5467 [hep-lat].
  %%CITATION = ARXIV:1310.5467;%%
  %4 citations counted in INSPIRE as of 14 Nov 2013

%\cite{Giudice:2013fza}
\bibitem{Giudice:2013fza}
  P.~Giudice, G.~Aarts, C.~Allton, A.~Amato, S.~Hands and
  J.~-I.~Skullerud,
  %``Electric charge susceptibility in 2+1 flavour QCD on an
  %anisotropic lattice,''
  PoS(LATTICE 2013) 492,
  arXiv:1309.6253 [hep-lat].
  %%CITATION = ARXIV:1309.6253;%%
  %3 citations counted in INSPIRE as of 14 Nov 2013

%\cite{Amato:2013oja}
\bibitem{Amato:2013oja}
  A.~Amato, G.~Aarts, C.~Allton, P.~Giudice, S.~Hands and
  J.~-I.~Skullerud,
  %``Transport coefficients of the QGP,''
  PoS(LATTICE 2013) 176,
  arXiv:1310.7466 [hep-lat].
  %%CITATION = ARXIV:1310.7466;%%
  %1 citations counted in INSPIRE as of 14 Nov 2013

%\cite{Evans:2013zca}
\bibitem{Evans:2013zca}
  P.~W.~M.~Evans, C.~Allton, P.~Giudice and J.~-I.~Skullerud,
  %``Charmonium Potentials At Non-Zero Temperature,''
  PoS(LATTICE 2013) 168,
  arXiv:1309.3415 [hep-lat].
  %%CITATION = ARXIV:1309.3415;%%
  %2 citations counted in INSPIRE as of 14 Nov 2013

%\cite{Harris:2013vfa}
\bibitem{Harris:2013vfa}
  T.~Harris, S.~M.~Ryan, G.~Aarts, C.~Allton, S.~Kim,
  M.~P.~Lombardo and J.~-I.~Skullerud,
  %``Bottomonium spectrum at finite temperature,''
  PoS(LATTICE 2013) 171,
  arXiv:1311.3208 [hep-lat].
  %%CITATION = ARXIV:1311.3208;%%

%\cite{Kelly:sewm2012}.
\bibitem{Kelly:sewm2012} A. Kelly, ``Spectral functions of charmonium
  in 2+1 flavour lattice QCD'', poster at Strong and Electroweak
  Matter 2012, Swansea, 10-13 July 2012.

%\cite{Lin:2008pr}
\bibitem{Lin:2008pr}
  H.~-W.~Lin {\it et al.}  [Hadron Spectrum Collaboration],
  %``First results from 2+1 dynamical quark flavors on an anisotropic
  %lattice: Light-hadron spectroscopy and setting the strange-quark
  %mass,''
  Phys.\ Rev.\ D {\bf 79} (2009) 034502
  [arXiv:0810.3588 [hep-lat]].
  %%CITATION = ARXIV:0810.3588;%%
  %150 citations counted in INSPIRE as of 14 Nov 2013
%\cite{Edwards:2008ja}

%\cite{Morningstar:2003gk}
\bibitem{Morningstar:2003gk}
  C.~Morningstar and M.~J.~Peardon,
  %``Analytic smearing of SU(3) link variables in lattice QCD,''
  Phys.\ Rev.\ D {\bf 69} (2004) 054501
  [hep-lat/0311018].
  %%CITATION = HEP-LAT/0311018;%%
  %270 citations counted in INSPIRE as of 14 Nov 2013

\bibitem{Edwards:2008ja}
  R.~G.~Edwards, B.~Joo and H.~-W.~Lin,
  %``Tuning for Three-flavors of Anisotropic Clover Fermions with
  %Stout-link Smearing,''
  Phys.\ Rev.\ D {\bf 78} (2008) 054501
  [arXiv:0803.3960 [hep-lat]].
  %%CITATION = ARXIV:0803.3960;%%
  %78 citations counted in INSPIRE as of 14 Nov 2013

%\cite{Edwards:2004sx}
\bibitem{Edwards:2004sx}
  R.~G.~Edwards {\it et al.}  [SciDAC and LHPC and UKQCD
    Collaborations],
  %``The Chroma software system for lattice QCD,''
  Nucl.\ Phys.\ Proc.\ Suppl.\  {\bf 140} (2005) 832
  [hep-lat/0409003].
  %%CITATION = HEP-LAT/0409003;%%
  %310 citations counted in INSPIRE as of 14 Nov 2013

\bibitem{bagel}
P.~A.~Boyle,
%``The BAGEL assembler generation library''.
Computer Physics Communications v180, 12 (2009) 2739-2748.
%http://www.ph.ed.ac.uk/~paboyle/bagel/Bagel.html

%\cite{Borsanyi:2012xf}
\bibitem{Borsanyi:2012xf}
  S.~Borsanyi, Y.~Delgado, S.~Durr, Z.~Fodor, S.~D.~Katz, S.~Krieg,
  T.~Lippert and D.~Nogradi {\it et al.},
  %``QCD thermodynamics with dynamical overlap fermions,''
  Phys.\ Lett.\ B {\bf 713} (2012) 342
  [arXiv:1204.4089 [hep-lat]].
  %%CITATION = ARXIV:1204.4089;%%
  %13 citations counted in INSPIRE as of 14 Nov 2013

%\cite{Amato:2013naa}
\bibitem{Amato:2013naa}
  A.~Amato, G.~Aarts, C.~Allton, P.~Giudice, S.~Hands and
  J.~-I.~Skullerud,
  %``Electrical conductivity of the quark-gluon plasma across the
  %deconfinement transition,''
  Phys.\ Rev.\ Lett.\  {\bf 111} (2013) 172001
  [arXiv:1307.6763 [hep-lat]].
  %%CITATION = ARXIV:1307.6763;%%
  %10 citations counted in INSPIRE as of 14 Nov 2013

\end{thebibliography}
\end{document}